\documentclass[prb,twocolumn,showpacs,preprintnumbers,amsmath,amssymb]{revtex4}

\usepackage{graphicx}
\usepackage{dcolumn}
\usepackage{bm}

\begin{document}


\title{Crystal nucleation of colloidal hard dumbbells}

\author{Ran Ni}
\email{r.ni@uu.nl}
\author{Marjolein Dijkstra}%
\email{m.dijkstra1@uu.nl}
\affiliation{%
Soft Condensed Matter, Debye Institute for NanoMaterials Science, Utrecht University, Princetonplein 5 3584CC,
Utrecht, The Netherlands
}%

\date{\today}

\begin{abstract}
Using computer simulations we investigate the homogeneous crystal nucleation in suspensions of colloidal hard dumbbells. The free energy
barriers are determined by Monte Carlo simulations using the umbrella sampling technique.  We calculate the
nucleation rates for the plastic crystal and the aperiodic crystal phase using the kinetic prefactor as determined from event driven molecular dynamics simulations. We find good agreement with the nucleation rates determined from spontaneous nucleation events observed in event driven molecular dynamics simulations within error bars of one order of magnitude. We study
the effect of aspect ratio of the dumbbells on the nucleation of plastic and aperiodic crystal phases and we also determine the structure of the
critical nuclei. Moreover, we find that the nucleation of the aligned CP1 crystal phase is strongly suppressed by a
high free energy barrier at low supersaturations and slow dynamics at high supersaturations.
\end{abstract}

\pacs{82.70.Dd, 64.60.Q-, 64.60.qe}
\keywords{nucleation, colloid, computer simulations, hard dumbbells, plastic crystal, aperiodic crystal}
\maketitle

\section{\label{sec:intro}Introduction}
Recent breakthroughs in particle synthesis produced a spectacular variety of anisotropic
building blocks.~\cite{glotzer2007} Colloidal particles with the shape of a dumbbell are one of the simplest anisotropic building blocks. Their unique morphologies
lead to novel self-organized structures. For instance, it was found that magnetic colloidal
dumbbells can form chain-like clusters with tunable chirality,~\cite{pine2008} while novel crystal structures have been predicted recently for asymmetric dumbbell particles consisting of a tangent large and small
hard sphere, which are
atomic analogs of NaCl, CsCl, $\gamma$CuTi, CrB, and $\alpha$IrV,
when we regard the two individual spheres of each dumbbell
independently.~\cite{filion2009}
Moreover, colloidal dumbbells also gain increasing scientific attention in recent years due to its potential
use in photonic applications. It has been shown that dumbbells on a face-centered-cubic
lattice where the spheres of the dumbbells form a diamond structure exhibit a complete
band gap~\cite{xia,lizy} while it is impossible to obtain a complete band gap in
systems consisting of spherical particles. A very recent calculation showed that for midrange aspect ratios, both
asymmetric and symmetric dumbbells have 2 - 3 large band gaps in the inverted lattice.~\cite{hosein}
Although these structures are not thermodynamically stable for  hard dumbbells,~\cite{smithline,mccoy,singer1990,vega1992a,vega1992b,vega1997,marechal2008} it does
show the promising potential of anisotropic particles in photonic applications.

New routes of synthesizing colloidal dumbbells make
it easy to control the aspect ratio.~\cite{carlos} In addition, by adding salt to the solvent, the interactions
between dumbbells can be tuned from long-ranged repulsive to hard interactions.
Although hard dumbbells were originally modeled for simple non-spherical diatomic molecules, such
as nitrogen, they are also a natural model system for studying the self-assembly of colloidal dumbbells.~\cite{mock2007,liddell2008,liddell2009,imhof2010}
The phase behavior of hard dumbbells has been extensively studied by density functional theory~\cite{smithline,mccoy} and
computer simulations.~\cite{singer1990,vega1992a,vega1992b,vega1997,marechal2008} The bulk phase diagram of hard dumbbells displays
three types of stable crystal structures.~\cite{smithline,mccoy,singer1990,vega1992a,vega1992b,vega1997,marechal2008} 
For small aspect ratio,
the dumbbells form a plastic crystal phase at low densities. The freezing into a cubic plastic crystal phase in which the dumbbells are positioned on a face-centered-cubic lattice but are free to rotate, has been determined using Monte Carlo simulations.~\cite{singer1990} These results have been refined by Vega, Paras, and Monson, who showed that at higher densities the cubic plastic crystal phase transforms into  an orientationally ordered crystal  CP1 phase. Additionally, these authors showed that the fluid-cubic plastic crystal coexistence region terminates at  $L/\sigma \simeq 0.38$,  where
$L$ is the distance between the centers of spheres and $\sigma$ is the diameter of the dumbbells. For longer dumbbells a fluid-CP1 coexistence region was found, whereas  the relative stability of the close-packed crystal structures CP1, CP2, and CP3, which only differ in the way the hexagonally packed dumbbell layers are stacked remained undetermined as the free energies are very similar.~\cite{vega1992a,vega1992b,vega1997} Moreover, these authors  showed by making an estimate for the degeneracy contribution to the free energy that dumbbells with $L/\sigma=1$ may form an aperiodic crystal phase.~\cite{vega1992a} The stability of such an aperiodic crystal structure in which both the orientations and
positions of the particles are disordered, while the spheres of each dumbbell are located on the lattice positions of a
random-hexagonal-close-packed (rhcp) lattice,  has been verified recently for $L/\sigma>0.88$.~\cite{marechal2008} In addition, it has been shown that the plastic crystal phase with the hexagonal-close-packed structure is more stable than the cubic plastic crystal for a large part of the stable plastic crystal region.~\cite{marechal2008} Although, the bulk phase diagram is well-studied, the kinetic pathways of the fluid-solid phase transitions are still unknown, and only a few studies have been devoted to  the crystal nucleation of anisotropic particles.~\cite{schilling2004,ran2010}
In the present work,  we investigate the nucleation of the plastic crystal phase of
hard dumbbells using computer simulations and study the effect
of aspect ratio of the dumbbells on the resulting nucleation rates and the structure and size of the critical nuclei. Moreover, for longer dumbbells we investigate crystal
nucleation of the  aperiodic crystal phase.
First,  we calculate the free energy barriers for nucleation using Monte Carlo (MC)
simulations with the umbrella sampling technique, which are then combined with  event driven molecular dynamics (EDMD) simulations to determine the kinetic prefactor and  the nucleation rates. Additionally, we determine the nucleation rates from spontaneous nucleation events observed in EDMD simulations. We compare the nucleation rates and critical nuclei obtained from
the umbrella sampling MC simulations with those from EDMD simulations.

The remainder of this paper is organized as follows. In Sec. II, we describe the methodology including the model
and simulation methods used. We present the results and discussions on the nucleation of
three types of crystal phases in suspensions of hard dumbbells in Sec. III. We end with some discussions and conclude in Sec. IV.

\section{\label{method}Methodology}
We consider a system of hard dumbbells consisting of two overlapping hard spheres with diameter $\sigma$
with the centers separated by a distance $L$. We define the aspect ratio as $L^*\equiv L/\sigma$, such that the model reduces to hard spheres for $L^*=0$ and to tangent spheres for $L^*=1$. We study crystal nucleation of hard dumbbells for $0 \leq L^* \leq 1$. We focus on the nucleation of the
plastic crystal phase ($0 \leq L^* < 0.4$) and the aperiodic crystal phase ($0.88 < L^* \leq 1$).~\cite{mccoy,vega1992a,vega1992b,vega1997,marechal2008}

\subsection{Order parameter}
In order to study the nucleation of the crystal phase, we require a cluster criterion that identifies the crystalline clusters in a metastable fluid.  In this work, we employ the order parameter
based on the local bond order parameter analysis of  Steinhardt {\it et al.}.~\cite{bop} We define for every particle
$i$, a $2l+1$-dimensional complex vector $\mathbf{q}_l(i)$ given by
\begin{equation}
	q_{lm}(i) = \frac{1}{N_b(i)}\sum_{j=1}^{N_b(i)}{\Upsilon_{lm}(\mathbf{\hat r}_{ij})}
\end{equation}
where $N_b(i)$ is the total number of neighboring particles of particle $i$, and $\Upsilon_{lm}(\mathbf{\hat r}_{ij})$
is the spherical harmonics for the normalized direction vector $\mathbf{\hat r}_{ij}$ between particle
$i$ and $j$, $l$ is a free integer parameter, and $m$ is an integer that runs from
$m=-l$ to $m= +l$. Neighbors of particle $i$ are defined as those particles which lie within a given cutoff radius $r_c$ from particle $i$.
In order to determine the correlation between the local environments of particle $i$ and $j$,  we define the rotationally invariant
function $d_l(i,j)$
\begin{equation}\label{d6}
	d_l(i,j) = \sum_{m=-l}^{l}{{\tilde q}_{lm}(i) \cdot {\tilde q}_{lm}^{*}(j)}
\end{equation}
where ${\tilde q}_{lm}(i)=q_{lm}(i)/\sqrt{\sum_{m = -l}^{l} |q_{lm}(i)}|^2$ and the asterisk is the complex conjugate .~\cite{tenwolde}
If $d_l(i,j) > d_c$, the bond between particle
(sphere) $i$  and $j$ is regarded to be solid-like or connected, where $d_c$ is the dot-product cutoff. We identify a particle (sphere) as  solid-like when it has at
least $\xi_c$ solid-like bonds.
We have chosen the symmetry index $l=6$ as the particles (spheres) display hexagonal order in the plastic crystal and the aperiodic crystal phase.
We have chosen $r_c=1.3 \sigma$, $d_c=0.7$, and $\xi_c=6$ in our simulations.  It has been shown recently that the choice of order parameter ($r_c$, $d_c$, and $\xi_c$)  does not affect the resulting
nucleation rate if it is not too restrictive.~\cite{filion2010,ran2010}

To analyze the structure of the critical nuclei, we use the averaged local bond order parameter $\overline{q}_l$ and
$\overline{w}_l$ proposed by Lechner and Dellago,~\cite{lechner} which allows us to identify each particle as fcc-like or hcp-like, provided the number of neighboring particles $N_b(i)\geq 10$:
\begin{equation}
	\overline{q}_l(i) = \sqrt{\frac{4\pi}{2l+1}\sum_{m=-l}^{l}{\left|\overline{q}_{lm}(i)\right|^2}}
\end{equation}
\begin{equation}
	\overline{w}_l(i) = \frac{\displaystyle
\sum_{m_1+m_2+m_3=0}\left(
\begin{array}{ccc}
l & l & l \\
m_1 & m_2 & m_3
\end{array}\right) \overline{q}_{lm_1}(i) \overline{q}_{lm_2}(i) \overline{q}_{lm_3}(i)
}{\displaystyle \left( \sum_{m=-l}^{l}\left|\overline{q}_{lm}(i)\right|^2 \right)^{3/2}}
\end{equation}
where
\begin{equation}
	\overline{q}_{lm}(i) = \frac{1}{N_b(i)+1}\sum_{k=0}^{N_b(i)}{q_{lm}(i)}
\end{equation}
The sum from $k=0$ to $N_b(i)$ runs over all neighbors of particle (sphere) $i$ plus the particle (sphere) $i$ itself.
While $q_{lm}(i)$ takes into account the structure of the first shell around particle $i$, the averaged $\overline{q}_{lm}(i)$, contains also the information of the structure of the second shell, which increases the accuracy of the crystal structure determination. In order to distinguish fcc-like and hcp-like particles, we employ $\overline{q}_4$ and
$\overline{w}_4$, as the order parameter distributions of pure fcc and hcp phases of Lennard-Jones and Gaussian core systems are well separated in the  $\overline{q}_4-\overline{w}_4$ plane.~\cite{lechner}

\subsection{Umbrella sampling}
The Gibbs free energy $\Delta G(n)$ for the formation of a crystalline cluster of size $n$ is  given by $\Delta G(n)/k_BT=\mathrm{const}-\ln[P(n)]$, where
$P(n)$ is the probability distribution function of finding a cluster of size $n$, $k_B$ is Boltzmann's constant, and $T$ the temperature.
As nucleation is a rare event and the probability to find a spontaneous nucleation event is very small in a brute force simulation within a reasonable time, one has to resort to specialized simulation techniques such as forward flux sampling, umbrella sampling or transition path sampling.  Here, we employ the method developed by Frenkel and
coworkers~\cite{auer2001} to calculate the free energy of the largest cluster. In this method, the sampling is  biased towards configurations
that contain clusters with a certain size. To this end, we introduce a biasing potential $\omega(\mathbf{r}^N)$, which is a harmonic function of the cluster size $n$:
\begin{equation}
	\beta \omega(\mathbf{r}^N) = \frac{1}{2}k\left[n(\mathbf{r}^N) - n_0\right]^2
\end{equation}
where $n(\mathbf{r}^N)$ is the size of largest cluster and $n_0$ is the center of the umbrella
sampling window whose width depends on $k$. In  this
work we set $k=0.2$.
By increasing the value of $n_0$, we
increase the size of the largest crystalline cluster in our system,
which enables us to cross the nucleation barrier.
If we define the average number of crystalline clusters with $n$ particles
by $\langle N_n \rangle$, one can calculate the probability distribution $P(n) = \langle N_n \rangle/N$ from which we can determine the Gibbs free energy $\Delta G(n)$.

\subsection{Event driven molecular dynamics simulations}
Since the potential between particles in systems of hard dumbbells is discontinuous, the pair interactions only change when particles collide. The particles perform elastic collisions when they encounter each other. We numerically identify and handle these collisions by using an EDMD simulation.~\cite{Donev2005737,Donev2005765}

Using MD simulations to determine the nucleation rate is straightforward. Starting with an equilibrated fluid configuration, an MD simulation is used to evolve the system until the largest cluster in the system exceeds the critical nucleus size. Then the nucleation rate is given by

\begin{equation}
  I = \frac{1}{\langle t \rangle V}
\end{equation}
where $\langle t \rangle$ is the averaged waiting time of forming a critical nucleus in a system of volume $V$.

\begin{figure}
 \includegraphics[width=0.4\textwidth]{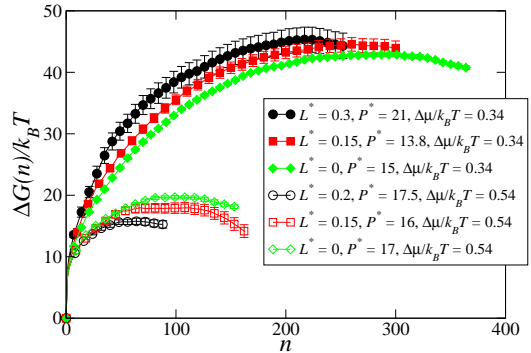}
 \caption{\label{fig1} Gibbs free energy $\Delta G(n)/k_B T$ as a function of cluster size $n$ for the nucleation of the
plastic crystal phase of hard dumbbells  with various aspect ratios $L^*=L/\sigma$ as displayed and supersaturation $\beta |\Delta \mu|=0.34$ (filled symbols) and 0.54 (open symbols).}
\end{figure}

\section{\label{results}Results and discussions}
In this section, we present the results on the nucleation of the plastic crystal, the aperiodic crystal and
the CP1 crystal phase in suspensions of hard dumbbells.

\begin{figure}
 \includegraphics[width=0.4\textwidth]{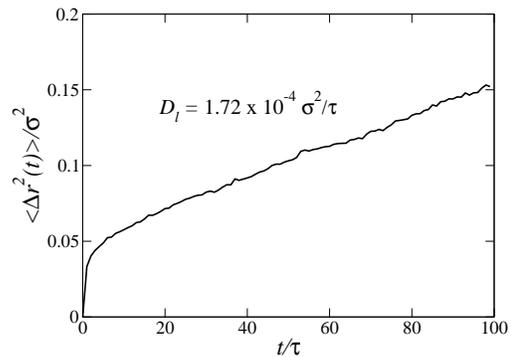}
\caption{\label{fig001} Mean square displacement $\langle \Delta r^2(t) \rangle$ as a function of time $t/\tau$ in a fluid of hard dumbbells with $L^*=0.3$ at $P^*=30$ (for $\beta |\Delta \mu|=0.47$).}
\end{figure}

\subsection{Nucleation of the plastic crystal phase}
We first investigate the nucleation of the plastic crystal phase of hard dumbbells.
Monte Carlo simulations with the umbrella sampling technique are performed on hard-dumbbell fluids with  $L^* = 0, 0.15$
and $0.3$ at supersaturation $\beta |\Delta \mu| = 0.34$ and with $L^*=0, 0.15$ and 0.2 for $\beta |\Delta \mu| = 0.54$
with $\beta=1/k_B T$. We have chosen a shorter aspect ratio for the highest supersaturation as the plastic crystal phase
for dumbbells with $L^* = 0.3$ becomes metastable with respect to the aligned CP1 phase for  $P^*=P\sigma^3/k_BT>30$,
i.e.,  $\beta |\Delta \mu| > 0.47$. The Gibbs free energy $\beta \Delta G(n)$ as a function of cluster size $n$ is  shown in
Fig.~\ref{fig1}.
 We clearly observe that at low supersaturation, i.e. $\beta |\Delta \mu| = 0.34$, the heights of the free energy barriers
increase slightly ($\sim 8\%$) with aspect ratio. More specifically, $\beta \Delta G^* = 42.9 \pm 0.3, 44.5 \pm 1.1$, and  $45.2 \pm 2$ for $L^* = 0, 0.15$
and $0.3$, respectively. According to classical nucleation theory (CNT), the nucleation barrier for a spherical nucleus
with radius $R$ is given by $\Delta G(R)=4 \pi \gamma R^2 - 4 \pi |\Delta \mu| \rho_s R^3/3$ with $\gamma$ the interfacial
tension, $|\Delta \mu|$ the chemical potential difference between the solid and fluid phase, and $\rho_s$ the bulk density of the
solid phase. CNT predicts a nucleation barrier height $\Delta G^* = (16 \pi/3)\gamma^3/(\rho_s |\Delta \mu|)^2$ and a critical
radius $R^* = 2 \gamma/\rho_s |\Delta \mu|$. The small increase in barrier height with aspect ratio can be
explained by the small increase in  the crystal-melt interfacial tensions that have been determined recently for the crystal planes (100), (110),
(111) using nonequilibrium work measurements with a cleaving procedure in MC simulations.~\cite{mu2006}  For a spherical
cluster, the surface tension is expected to be an average over the crystal planes, i.e.,  $\beta \gamma d^2 =0.58, 0.57,$
and 0.60, for $L^* = 0, 0.15$
and $0.3$, respectively, where $d^3 = \sigma^3(1+3/2 L^*-1/5 L^*\mbox{}^3)$. Another work by Davidchack {\em et al.} found a slightly lower value for the averaged interfacial tension
of hard spheres, i.e, $\beta \gamma d^2 =0.559$.~\cite{davidchack2006} Using $\beta\Delta G^*=42.9$ and the more precise value for the surface tension $\beta\gamma d^2=0.559$, and the values for $\beta \gamma d^2$ and the bulk density $\rho_s$  for varying $L^*$ presented in Table~\ref{tab1}, CNT predicts a slightly larger increase in barrier height upon increasing $L^*$, i.e., $\beta \Delta G^*=45.6$ and 50.49, for $L^*=0.15$ and 0.3, respectively. However, when
the supersaturation is increased to $\beta |\Delta \mu| = 0.54$, we find a decrease in barrier height upon increasing the aspect ratio as shown in Fig.~\ref{fig1} and Table~\ref{tab1}, which cannot be explained by CNT. Apparently, the pressure dependence of the surface tension is different for dumbbells with various aspect ratios.

\begin{figure*}
 \includegraphics[width=0.3\textwidth]{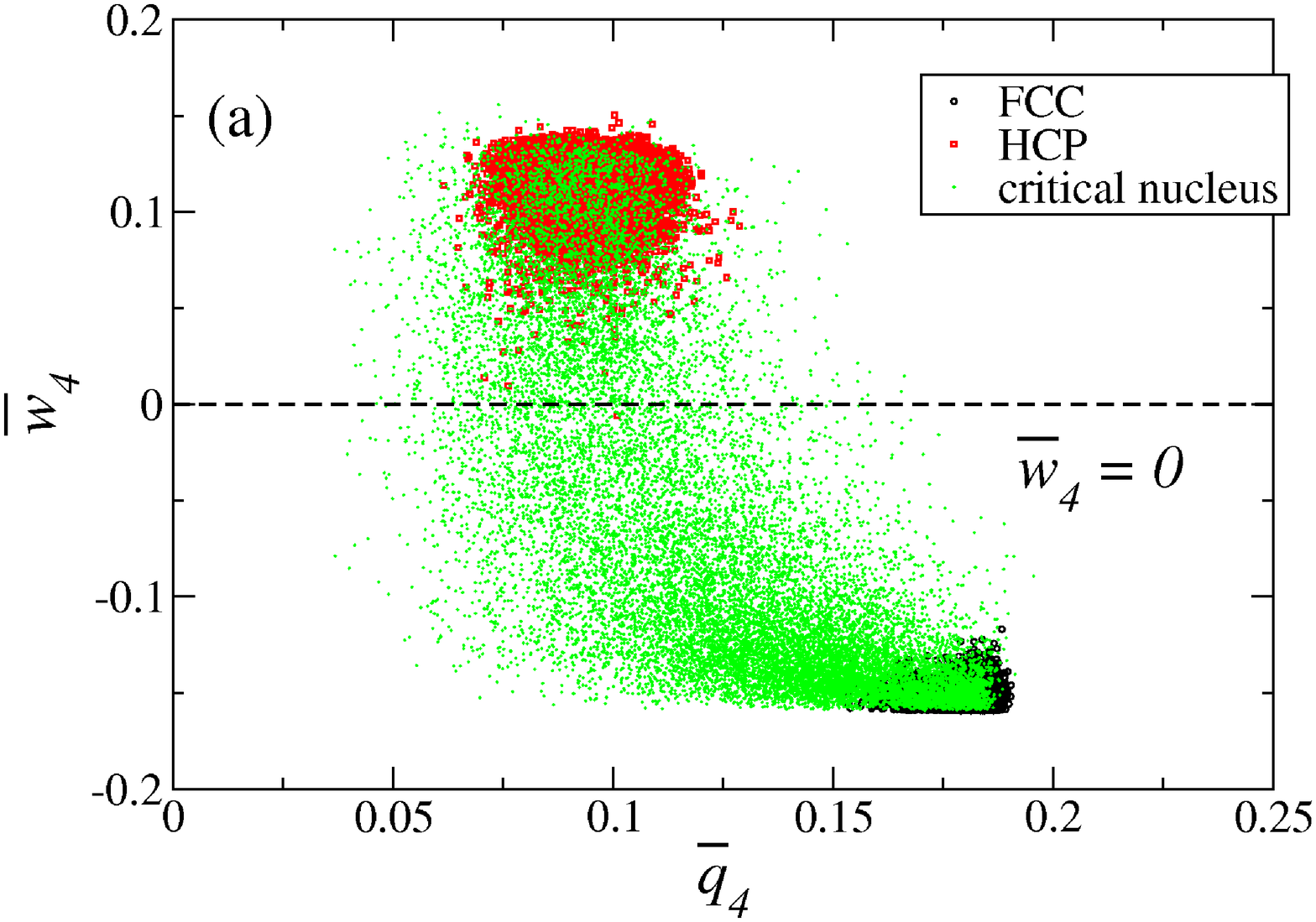}
 \includegraphics[width=0.3\textwidth]{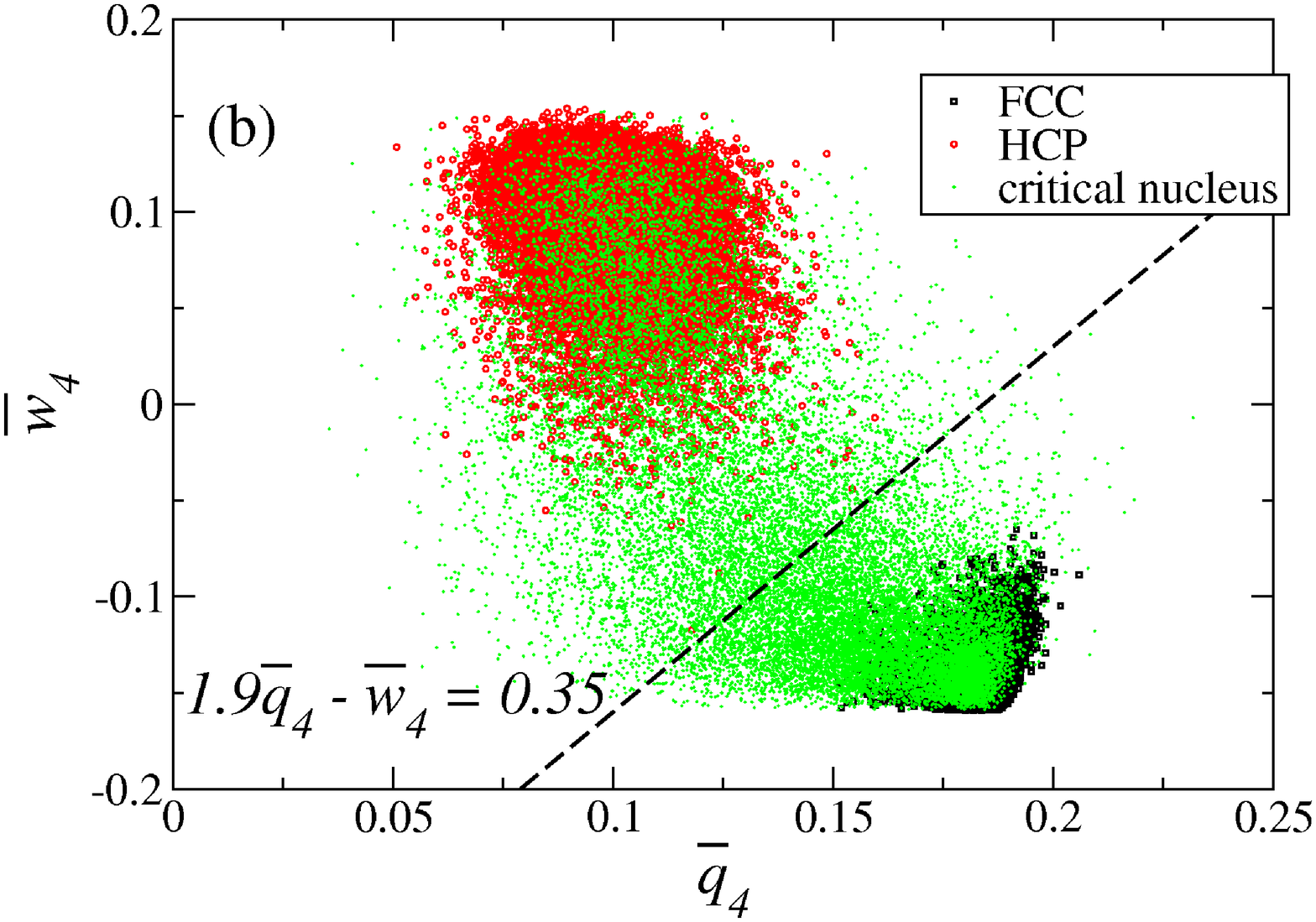}
 \includegraphics[width=0.3\textwidth]{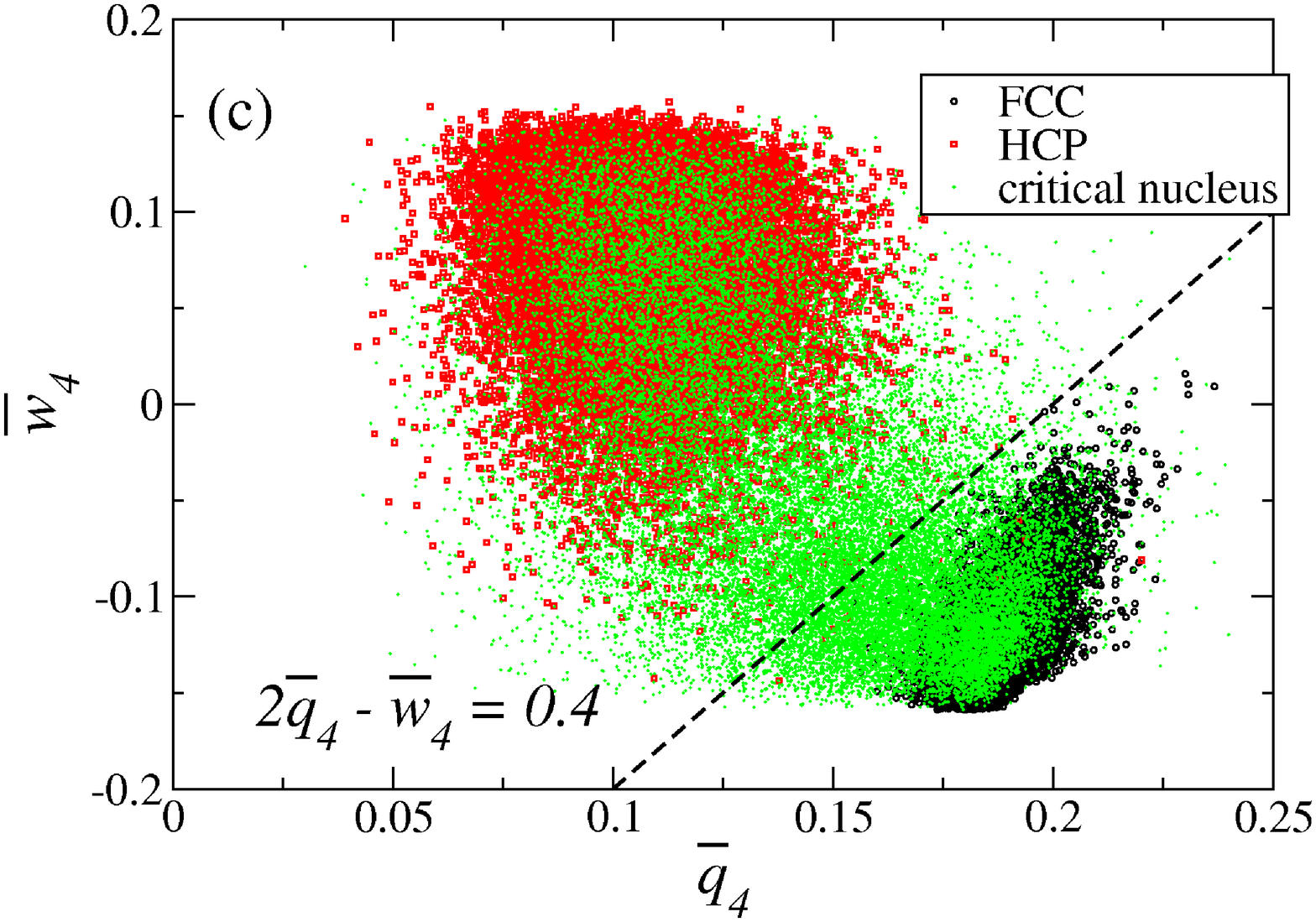}
 \caption{\label{fig22} Distribution of particles in the critical nuclei  for the plastic crystal nucleation in hard dumbbell systems with $L^*=0$ (a), 0.15 (b), and 0.3 (c) as obtained from umbrella sampling MC simulations at a supersaturation $\beta |\Delta \mu|=0.34$
in the $\overline{q}_4-\overline{w}_4$ plane compared with those for pure fcc and hcp plastic crystal phases with corresponding
pressures. The dashed lines are used to
distinguish the fcc-like and hcp-like particles, and the formulas are next to them.}
\end{figure*}

The nucleation barriers obtained from umbrella sampling MC simulations can also be used to determine the nucleation rates as given by~\cite{auer2001}:
\begin{equation}
 I = \kappa \exp\left(-\beta \Delta G^*\right)
\end{equation}
where $\kappa$ is the kinetic prefactor given by
$\kappa = \rho_l f_{n^*}~\sqrt{|\Delta G^{''}(n^*)|/2\pi k_BT}$,  $\rho_l$ is the number density of particles in
the fluid phase, $f_{n^*}$ the rate at which particles are attached to the critical nucleus, $\Delta G^{''}(n^*)$ is the second derivative on the top of the Gibbs free energy barrier.
The attachment rate can be  calculated from the mean square deviation of the cluster size at the
top of the free energy barrier by
\begin{equation}
  f_{n^*} = \frac{1}{2} \frac{\langle [n(t)-n(0)]^2 \rangle}{t}
\end{equation}
where $n(t)$ is the cluster size at time $t$. The mean square deviation of the cluster size can be determined from  EDMD  simulations starting from  configurations
 at the top of the free energy barriers. Using the results for the attachment rates and the nucleation barriers obtained from umbrella sampling MC simulations, we can determine the nucleation rates, which we compare with those obtained directly from spontaneous nucleation events in EDMD simulations.
We observed a large variance in the attachment rates calculated for different nuclei. We used 10 independent configurations on
the top of the barrier and followed 10 trajectories for each of them to determine the attachment rates. Taking into account the statistical errors in the free energy barriers and attachment rates, we estimate that the error in the resulting nucleation rates is one order of magnitude.
 In order to exclude the effect of dynamics, we
compare the nucleation rates for the plastic crystal phase in long-time diffusion times, i.e. $\tau_L = \sigma^2/6 D_l$ with $D_l$ the long-time diffusion coefficient. We calculate $D_l$ by measuring the mean square displacement at supersaturation $\beta |\Delta \mu| = 0.34$ and 0.54 as shown in Table~\ref{tab1} for various aspect ratios. We clearly observe that the dynamics becomes slower for increasing aspect ratio $L^*$, resulting in  long-time diffusion coefficients $D_l \tau/\sigma^2= 0.012, 0.01, 0.0023$ for $L^* = 0,0.15$ and $0.3$ at $\beta |\Delta \mu| = 0.34$ with $\tau = \sigma \sqrt{m/k_BT}$. At higher supersaturation $\beta |\Delta \mu| = 0.54$, we find even smaller values for $D_l$, i.e., $D_l \tau/\sigma^2= 0.0078, 0.006, 0.003$ for $L^* = 0,0.15$ and $0.2$, respectively.

\begin{figure*}
 \includegraphics[width=0.8\textwidth]{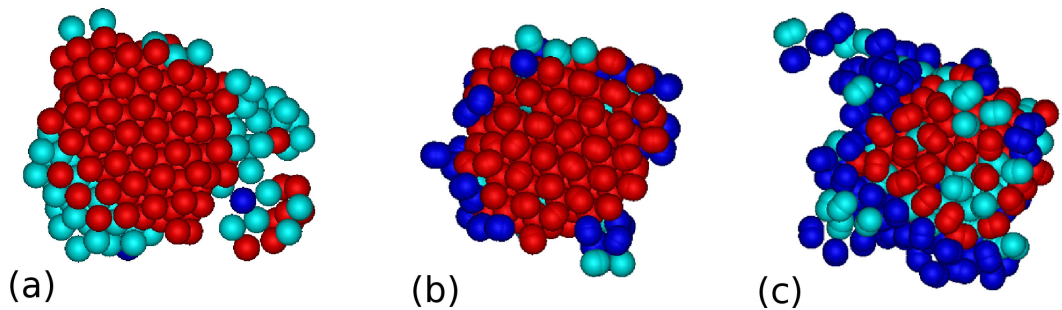}
 \caption{\label{fig3} Typical configurations of critical nuclei for the plastic crystal nucleation of hard dumbbells with aspect ratios $L^*=0$ (a), 0.15 (b), and 0.3 (c) at supersaturation
$|\Delta \mu| = 0.34k_BT$. The red (dark grey) particles are fcc-like, the blue particles are hcp-like particles, while the light blue (light grey) particles are undetermined.}
\end{figure*}

The resulting nucleation rates in  units of the long-time diffusion coefficient are shown in Table~\ref{tab1}.
We wish to make a few remarks here. First,  the nucleation rates obtained from spontaneous nucleation events observed in EDMD simulations  agree well with the ones
obtained from umbrella sampling  MC simulations  within error bars of one order of magnitude, which means that the nucleation results obtained
from the umbrella sampling MC simulations are reliable.  Secondly, we clearly observe that  the nucleation rates
for the different aspect ratios ranging from $L^* = 0$ to $0.3$ are remarkably similar as  the differences are within the errorbars for both supersaturations.

Finally, we made an attempt to study spontaneous nucleation of dumbbells with $L^*=0.3$ at supersaturation
$\beta|\Delta \mu| = 0.54$ using event-driven MD simulations. As already mentioned above,  the plastic
crystal phase for dumbbells with $L^* = 0.3$  becomes metastable with respect to the aligned CP1 phase for  $P^*=P\sigma^3/k_BT>30$, i.e.,  $\beta |\Delta \mu| > 0.47$.
Hence, we would expect to find the nucleation of the CP1 phase here. However, we find that the nucleation
is severely hampered due to slow dynamics, which can be appreciated from Fig.~\ref{fig001}, where we plot
the mean square displacement for  $\beta |\Delta \mu| = 0.47$. The resulting  long-time diffusion
coefficient $D_l = 1.72\times 10^{-4}\sigma^2/\tau$ is at least one order of magnitude smaller than the long-time diffusion coefficients at $\beta|\Delta\mu|=0.54$, where we observed spontaneous nucleation for $L^*=0, 0.5$, and 0.2.

\begin{table*}\caption{\label{tab1}
Nucleation rates $I\sigma^5/6D_l$ for the nucleation of the plastic crystal phase in systems of hard dumbbells with elongation $L^*$, at pressure $P\sigma^3/k_B T$, and supersaturation $\beta|\Delta \mu|$. $\rho_s d^3$ is the number density of dumbbells in the solid phase, $\beta \Delta G(n^*)$ is the barrier height, and $|\beta \Delta G^{''}(n^*)|$ is the second derivative of the Gibbs free energy  at the critical nucleus size $n^*$, i.e., the number of dumbbells in the critical cluster. $f_{n^*}/6D_l$ is the attachment rate in units of the long-time diffusion coefficient $D_l$.}
\begin{ruledtabular}
\begin{tabular}{ccccccccccc}
  $L^*$	& $P\sigma^3/k_BT$ & $\beta |\Delta \mu|$ & $\rho_sd^3$\footnotemark[1] & $n^*$  & $\beta \Delta G(n^*)$ & $|\beta \Delta G^{''}(n^*)|$
& $f_{n^*}/6D_l$ & $D_l\tau/\sigma^2$ & $I\sigma^5/6D_l$ (US) & $I\sigma^5/6D_l$ (MD)\\
0 & 15 & 0.34 & 1.107 & 300 & $42.9\pm0.3$ & $5.1\times 10^{-4}$ & 4550 & $0.012$ & $9.6\times10^{-18\pm1}$ & -\\
0.15 & 13.8 & 0.34 & 1.104 & 265 & $44.5\pm1.1$ & $6.0\times 10^{-4}$ & 3700 & $0.01$ & $1.4\times10^{-18\pm1}$ & -\\
0.3 & 21 & 0.34 & 1.163 & 220 & $45.2\pm2$ & $1.0\times 10^{-3}$ & 7464 & $0.0023$ & $1.7\times10^{-18\pm1}$ & -\\
0 & 17 & 0.54 & 1.136 & 102 & $19.6\pm0.3$ & $1.2\times 10^{-3}$ & 3980 & $0.0078$ & $1.7\times10^{-7\pm1}$ & $1.6\times 10^{-7}$\footnotemark[2]\\
0.15 & 16 & 0.54 & 1.131 & 70 & $18.0\pm0.7$ & $9.7\times 10^{-4}$ & 3779 & $0.006$ & $6.1\times10^{-7\pm1}$ & $3.5\times 10^{-7\pm1}$\\
0.2 & 17.5 & 0.54 & 1.143 & 65 & $15.8\pm0.5$ & $2.0\times 10^{-3}$ & 2682 & $0.003$ & $5.5\times10^{-6\pm1}$ & $4.4\times 10^{-6\pm1}$\\
\end{tabular}
\end{ruledtabular}
\begin{flushleft} \footnotemark[1]{$d^3=\sigma^3(1+3/2L^*-1/2L^{*3})$~\cite{mu2006}}\\
\footnotemark[2]{Extrapolated from Ref.~\onlinecite{filion2010}}
\end{flushleft}
\end{table*}

In umbrella sampling MC simulations, we can ``fix'' the simulations at the top of the nucleation barrier which allows us to
study the properties of the critical nuclei. We investigate the effect of the particle anisotropy on the structure of the critical nuclei using the  order parameters $\overline q_4$ and $\overline w_4$ as defined above.
At supersaturation $\beta |\Delta \mu| = 0.34$, the size of the critical nuclei is $n \simeq 250$ which is sufficiently large to determine the crystal structure of the  nuclei. For each dumbbell, we calculate  the averaged local bond order parameter $\overline q_4$ and $\overline w_4$, provided the particle has $N_b(i)\ge10$ neighbors.
The distribution of particles in the critical nuclei are presented as scatter plots in the  $\overline q_4-\overline w_4$ plane along with those for
pure fcc and hcp plastic crystal phases of dumbbells with $L^*=0, 0.15$, and 0.3, at corresponding pressures. From Fig. \ref{fig22}, we clearly observe that the critical nuclei for $L^*=0$ and 0.15, contains predominantly  fcc-like rather than hcp-like particles. In order to distinguish the fcc-like and hcp-like particles more quantitatively, we divide the $\overline q_4-\overline w_4$ plane by a straight line in such a way that the particle distributions for the pure fcc and hcp plastic crystal phases are maximally separated. We plot the  criteria to distinguish  fcc-like and hcp-like particles as dashed straight lines in Fig. \ref{fig22} with the corresponding formula. We note, however, that the criteria seem to be arbitrarily chosen, but the identification of fcc-like and hcp-like particles for typical nuclei seems to be less sensitive on the precise details of these criteria.
Typical snapshots of the critical nuclei for  $L^*=0, 0.15$, and 0.3 are shown in
Fig.~\ref{fig3}, where the color-coding denotes the identity (fcc-like, hcp-like or undetermined) of the particle using these criteria. As we did not calculate the averaged local bond order parameter $\overline q_4$ and $\overline w_4$ for particles with $N_b(i)<10$ neighbors,  the identity of these particles remains  undetermined.  We clearly observe that the critical nuclei for  $L^*=0, 0.15$ contains mainly fcc-like particles. The particle distributions becomes broader  for the pure fcc and hcp plastic crystal phases upon increasing $L^*$ and consequently it becomes more difficult to distinguish fcc-like and hcp-like particles. However, the fraction of hcp-like particles seems to increase with increasing particle elongation.  This agrees with the results from free energy calculations of hard dumbbell systems, where it has been shown that the hcp plastic crystal phase is more stable than the one with an fcc structure at $L^*\ge 0.15$.~\cite{marechal2008} It is worth noting here that recent nucleation studies of  hard spheres showed that the critical nuclei contain approximately 80\% fcc-like particles.~\cite{filion2010} As the free energy difference per particle between bulk fcc and  hcp phases is only about 0.001 $k_B T$ at melting, the predominance for fcc-like particles is attributed to surface effects.

\begin{figure}
 \includegraphics[width=0.4\textwidth]{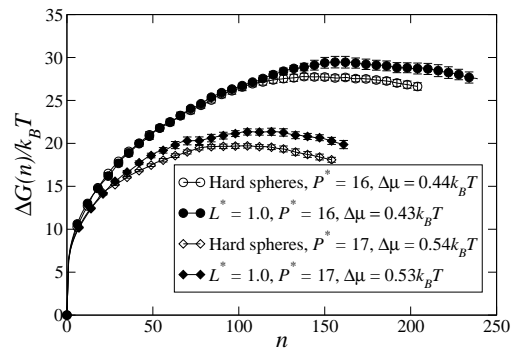}
 \caption{\label{fig5} Gibbs free energy  $\Delta G(n)$  as a function of number of spheres $n$  in the largest cluster for the nucleation of the (aperiodic) crystal phase of hard dumbbells and of hard spheres.}
\end{figure}

\begin{figure}
 \includegraphics[width=0.4\textwidth]{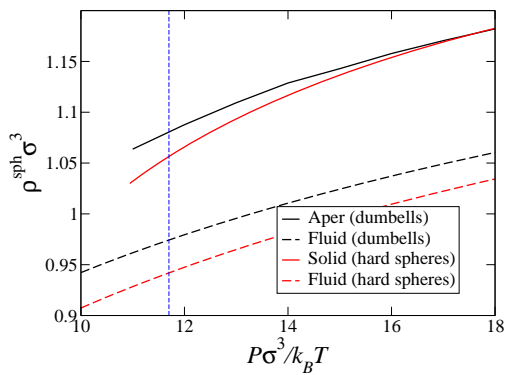}
 \caption{\label{fig31} Equation of state (EOS, i.e., $\beta P \sigma^3$ vs the number density of spheres $\rho^{sph} \sigma^3$  for a system of hard spheres and hard dumbbells with $L^*=1.0$. For the fluid and solid phase of  hard
spheres, the Carnahan-Starling~\cite{CS1969} and Speedy~\cite{speedy} EOS are plotted. The EOS of hard dumbbells for the
fluid phase is obtained from Ref.~\onlinecite{eosdb}. The dashed vertical line denotes the bulk coexistence pressure  of hard dumbbells
with $L^*=1.0$.}
\end{figure}

\subsection{Nucleation of the aperiodic crystal phase}
For more elongated dumbbells, i.e. $L^* > 0.88$, the orientationally disordered aperiodic crystal phase becomes stable,~\cite{vega1992a,vega1992b,vega1997,marechal2008} in which the individual spheres of the dumbbells are  on  a random hcp lattice whereas
the orientations of the dumbbells are random.  In this section, we investigate the nucleation
of the aperiodic crystal phase of hard dumbbells with different aspect ratios.
We perform Monte Carlo simulations using the umbrella sampling technique to determine the Gibbs
free energy as a function of cluster size for hard dumbbells  with $L^*=1.0$ and  supersaturation $P^* = 16$ and
17. The order parameter that is employed here in the umbrella sampling technique is equal to the number of spheres $n$ (and thus not the number of dumbbells) in the largest crystalline cluster in the system. Thus, we check for each individual sphere whether or not it belongs  to the largest crystalline cluster, and as a consequence, the whole dumbbell can be part of the largest cluster or only one sphere of the dumbbell can belong to the cluster, or the whole dumbbell is regarded to be fluid-like. Consequently, it is convenient to introduce a bulk chemical potential per sphere, which equals 0.5 times the bulk chemical potential per dumbbell $\mu^{sph}=\mu/2$.  We compare the results with those for hard spheres at the same pressure in Fig.~\ref{fig5}.  Since the bulk
pressure for the solid-fluid transition of hard dumbbells with $L^*=1$ is remarkably close to that of hard spheres
$\beta P_{coex} \sigma^3 = 11.8$,~\cite{vega1992a,vega1992b,vega1997,marechal2008} one might naively expect that the nucleation barriers should be compared at the
same dimensionless pressure.   However, we observe that at the same  pressure, the nucleation barrier for the aperiodic crystal
phase of hard dumbbells is slightly higher than that of hard spheres.  CNT predicts that
the barrier height is given by $\Delta G^*= (16 \pi/3)\gamma^3/(\rho_s^{sph} |\Delta \mu^{sph}|)^2$, and hence a difference
in barrier height should be due to a difference in the interfacial tension $\gamma$, the density of spheres in the solid phase
$\rho_s^{sph}$, or in $|\Delta \mu^{sph}|$. As the reduced density of spheres $\rho_s^{sph} \sigma^3$ in the aperiodic crystal phase is very close to that of a solid phase of hard spheres
 at $P^*=16$ and 17, and the interfacial tensions $\beta \gamma \sigma^2$ are also expected to be very similar, the
difference in barrier height  can only be caused by a difference in $|\Delta \mu^{sph}|$.  We therefore calculated  more
accurately the bulk chemical potential difference per sphere between the solid and the fluid phase  using
\begin{equation}\label{eq1}
|\Delta \mu^{sph}| = \int_{P_{coex}}^{P} {\left( \frac{1}{\rho_l^{sph}} - \frac{1}{\rho_s^{sph}}\right) \mathrm{d}P  }
\end{equation}
where $\rho_l^{sph}$ and $\rho_s^{sph}$ are the density of spheres in the liquid and solid phase. In Fig.~\ref{fig31}, we plot the
equation of state for the fluid and solid phase of hard spheres from Ref.~\onlinecite{CS1969,speedy}
along with the equation of state for the fluid phase of hard dumbbells for $L^*=1$ from Ref.~\onlinecite{eosdb}.
In addition, we determined the equation of state for the solid phase using EDMD simulations.
Using these results  and Eq. \ref{eq1}, we indeed find that the supersaturation
$\beta |\Delta \mu^{sph}|$ per sphere is $\sim 2.3\%$ smaller for hard dumbbells than for hard spheres, resulting in an increase
in barrier height of $\sim 5\%$, which perfectly matches our results. We conclude that the difference in the
height of the nucleation barrier between the aperiodic crystal phase of dumbbells with $L^*=1.0$ and the
hard-sphere crystal is mostly due to the difference in $|\Delta \mu^{sph}|$.

\begin{figure}
 \includegraphics[width=0.4\textwidth]{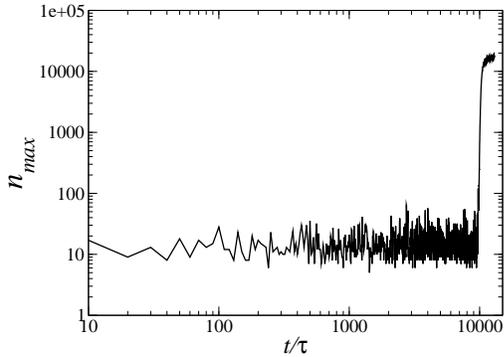}
 \caption{\label{fig51} Number of spheres in the largest cluster $n_{max}$ as a function of time $t/\tau$ for a typical trajectory obtained from EDMD simulations for the aperiodic crystal nucleation of hard dumbbells with $L^*=1.0$,
$N=16 000$ and $P^*=17$.}
\end{figure}

\begin{figure}
 \includegraphics[width=0.4\textwidth]{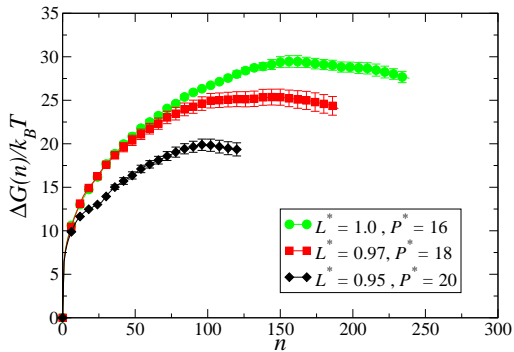}
 \caption{\label{fig52} Gibbs free energy $\Delta G(n)$ as a function of the number of spheres $n$ in the largest crystalline  cluster  for the aperiodic crystal nucleation of hard dumbbells with $L^*=0.95, 0.97$, and 1
at supersaturation $\beta|\Delta \mu^{sph}| = 0.43$.}
\end{figure}

Moreover, we also performed EDMD simulations for the spontaneous nucleation of the aperiodic crystal phase of dumbbells at $P^*=17$ in system of $N=16 000$
hard dumbbells. The number of spheres in the biggest cluster as a function of time from a typical MD simulation is shown in Fig.~\ref{fig51}.
We find that the size of critical nuclei in spontaneous nucleation is around $100$ spheres which agrees well with the result obtained
from umbrella sampling MC simulations shown in Fig.~\ref{fig5}. The nucleation rate obtained from spontaneous nucleation events observed in MD simulations  is
$I \sigma^5/6D_l = 7.3\times 10^{-8\pm1}$ which agrees very well with the rate obtained from umbrella sampling MC simulations,
$I \sigma^5/6D_l = 2.8\times 10^{-8\pm1}$, within the error bars of one order of magnitude.

\begin{figure}
 \includegraphics[width=0.4\textwidth]{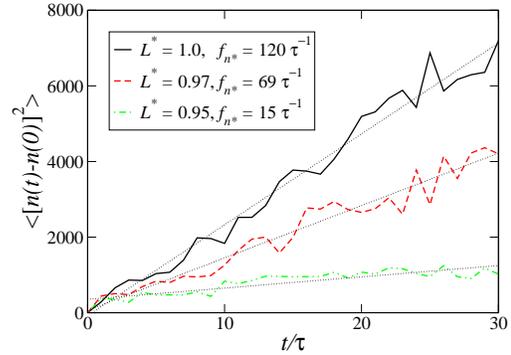}
 \caption{\label{fig53}
Mean square deviation of the cluster size $<\lbrack n(t)-n(0)\rbrack^2 >$ as a function of time $t/\tau$ for hard dumbbells with $L^*=0.95, 0.97,$ and $1.0$
at supersaturation $\beta|\Delta \mu^{sph} |= 0.43$. The resulting attachment rates $f_{n^*}$  are listed in units of $\tau^{-1}$.}
\end{figure}

\begin{figure}
 \includegraphics[width=0.4\textwidth]{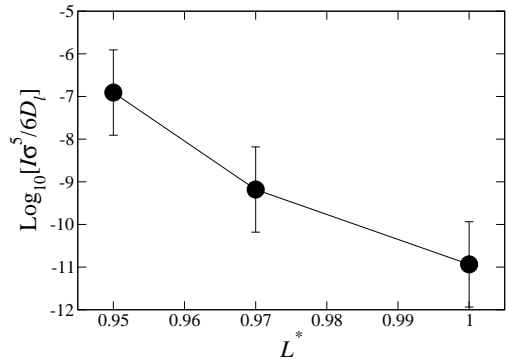}
 \caption{\label{fig54} Nucleation rate $I\sigma^5/6D_l$ for the aperiodic crystal phase as a function of the aspect ratio $L^*$ of hard dumbbells  at supersaturation $\beta |\Delta \mu^{sph}| = 0.43$.}
\end{figure}

Furthermore, we study the effect of aspect ratio on the nucleation of the  aperiodic crystal phase, and the free energy barriers for hard dumbbells
with aspect ratios $L^*=0.95, 0.97,$ and $1.0$ at  supersaturation $\beta|\Delta \mu^{sph}|=0.43$. We plot $\Delta G(n)$ as a function of cluster size $n$, i.e., the number of spheres in the cluster,  in Fig.~\ref{fig52}. We observe that
at the same supersaturation the barrier
height decreases upon decreasing the elongation of the dumbbells. According to classical nucleation theory,  $\Delta G^* \propto \gamma^3/(\rho^{sph}_s |\Delta \mu^{sph}|)^2$, where $\Delta \mu^{sph}$ is the supersaturation per sphere with $\rho_s^{sph}$ the bulk density of spheres in the solid phase. As shown in Table~\ref{tab2}, $\rho_s^{sph} \sigma^3$ is very similar for  $L^*=0.95, 0.97,$ and $1.0$, and
we argue that the interfacial tension of the aperiodic crystal decreases upon decreasing the elongation of the dumbbells. In order to calculate the nucleation rates,
we perform EDMD simulations starting from  configurations on the top of the free energy barriers. We plot the mean square deviation of the cluster
size as a function of time  in Fig.~\ref{fig53}. We find that the attachment rate decreases significantly as the
anisotropy of the dumbbells decreases. The resulting nucleation rates in units of the long time diffusion coefficient are shown in Fig.~\ref{fig54}. We clearly observe that
 at fixed supersaturation the nucleation rate increases with decreasing dumbbell elongation. However, in the phase diagram of hard dumbbells,~\cite{vega1992a,vega1992b,vega1997,marechal2008} the pressure range where the aperiodic crystal phase is thermodynamically
stable shrinks significantly when the aspect ratio decreases. As a result, it is not possible to increase the supersaturation further for shorter dumbbells, although the nucleation rates are already much higher for shorter ones than for longer ones at the same supersaturation.

\begin{figure}
 \includegraphics[width=0.45\textwidth]{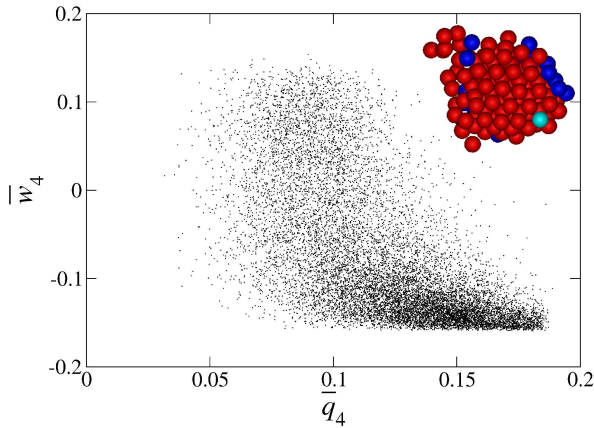}
 \caption{\label{fig55} Distribution of the spheres in the critical nuclei as obtained from umbrella sampling MC simulations
in $\overline{q}_4-\overline{w}_4$ plane in systems of hard dumbbells with $L^*=1.0$
at  supersaturation $\beta|\Delta \mu^{sph}| = 0.43$. Inset: Typical configuration of a critical nucleus.
Red denotes fcc-like spheres and blue denotes hcp-like spheres while the
light blue are the undetermined ones.}
\end{figure}

Additionally, we also study the structure of the critical nuclei by calculating the averaged local bond order parameter $\overline q_4$ and $\overline w_4$, provided the sphere has $N_b(i)\ge10$ neighbors.  The distribution of spheres in the critical nuclei are presented as scatter plots in the
$\overline{q}_4-\overline{w}_4$ plane in Fig.~\ref{fig55} for  $L^*=1.0$. We observe only a  few spheres with $\overline{w}_4>0$ and $\overline{q}_4< 0.1$, as most of the spheres are in the area
of $\overline{w}_4<0$ and $\overline{q}_4> 0.1$, which is very similar to the scatter plots for  hard spheres shown in Fig.~\ref{fig22}a. Consequently, the  critical nucleus of the aperiodic crystal phase of hard dumbbells contains also  more fcc-like than hcp-like particles, similar to the critical nuclei observed in hard-sphere nucleation.~\cite{filion2010}
 A typical configuration of a critical nucleus is shown in the inset of Fig.~\ref{fig55}, where the spheres are considered to be  fcc-like if $\overline{w}_4<0$.

\begin{table*}\caption{\label{tab2}Nucleation rates $I\sigma^5/6D_l$ for the nucleation of the aperiodic crystal phase in systems of hard dumbbells with elongation $L^*$, at pressure $P\sigma^3/k_B T$, and supersaturation per sphere $\beta|\Delta \mu^{sph}|$. $\rho^{sph}_s\sigma^3$ is the number density of spheres in the solid phase, $\beta \Delta G(n^*)$ is the barrier height, and $|\beta \Delta G^{''}(n^*)|$ is the second derivative of the Gibbs free energy  at the critical nucleus size $n^*$, i.e., the number of spheres in the critical cluster. $f_{n^*}/6D_l$ is the attachment rate in units of the long-time diffusion coefficient $D_l$.}
\begin{ruledtabular}
\begin{tabular}{ccccccccccc}
  $L^*$	& $P\sigma^3/k_BT$ & $\beta |\Delta \mu^{sph}|$ & $\rho^{sph}_s\sigma^3$ & $n^*$  & $\beta \Delta G(n^*)$ & $|\beta \Delta G^{''}(n^*)|$
& $f_{n^*}/6D_l$ & $D_l\tau/\sigma^2$ & $I\sigma^5/6D_l$ (US) & $I\sigma^5/6D_l$ (MD)\\
1 & 17 & 0.53 & 1.170 & 115 & $21.4\pm0.4$ & $1.2\times 10^{-3}$ &  2813 & 0.0026 & $2.0\times 10^{-8\pm1}$  & $7.3\times10^{-8\pm 1}$\\
1 & 16 & 0.43 & 1.158 & 170 & $29.5\pm0.6$ & $9.4\times 10^{-4}$ & 5556  & 0.0036 & $1.1\times 10^{-11\pm1}$ & - \\
0.97 & 18 & 0.43 & 1.171 & 140 & $25.3\pm0.9$ & $8.4\times10^{-4}$ & 5228  & 0.0022 & $6.6\times 10^{-10\pm1}$ & - \\
0.95 & 20 & 0.43 & 1.182 & 100 & $19.9\pm0.7$ & $3.0\times10^{-3}$ & 2273  & 0.0011 & $1.2\times 10^{-7\pm1}$ & - \\

\end{tabular}
\end{ruledtabular}
\end{table*}

\subsection{Slow dynamics of hard dumbbells}
\begin{figure}
 \includegraphics[width=0.45\textwidth]{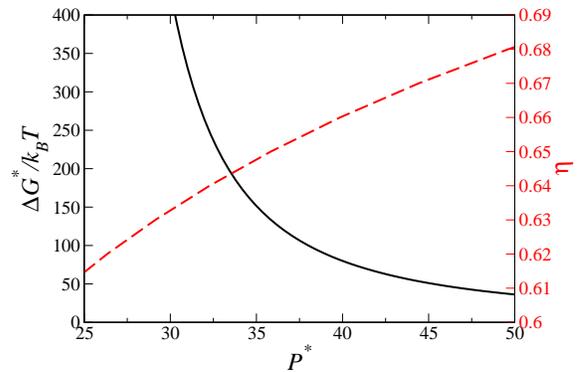}
 \caption{\label{fig6} Estimated height of the Gibbs free energy barrier $\Delta G^*/k_B T$ obtained from classical nucleation theory (solid line) and the packing fraction $\eta$
in the supersaturated fluid phase~\cite{eosdb} (dashed line) as a function of pressure $P^*$ for the nucleation of the CP1 phase of hard dumbbells with $L^*=0.4$.}
\end{figure}

The phase diagram of hard dumbbells shows a stable aligned CP1 crystal phase  at infinite pressure for all aspect ratios of the dumbbells, and a fluid-CP1 coexistence region for  $0.4 \le L^* \le 0.8$.~\cite{vega1992a,vega1992b,vega1997,marechal2008} The surface tension for the fluid-CP1 interface  of hard dumbbells with
$L^*=0.4$ is  $\beta \gamma \sigma^2 \simeq 1.8$.~\cite{mu2006} The height of free energy barrier is given by $\Delta G^* = 16\pi\gamma^3/3(\rho_s |\Delta \mu|)^2$ in CNT. If we assume that the interfacial tension does not change significantly with increasing pressure,  we can estimate
the free energy barrier height as a function of pressure by integrating the Gibbs-Duhem equation to obtain  $|\Delta \mu|$. The barrier height $\Delta G^*$ and the packing fraction $\eta$ for the fluid phase are shown in Fig.~\ref{fig6} as a function of the pressure $P^*$. We find that
the barrier height  $\Delta G^*$ is extremely high, and  only  becomes less than $50k_BT$ for $P^*>45$, corresponding to a
packing fraction of the fluid phase $\eta>0.67$. However, if the interfacial tension  increases with increasing pressure as shown in Ref.~\onlinecite{auer2001surf}, the
``actual'' height of free energy barrier can become even higher. As a consequence nucleation of the CP1 crystal phase is an extremely rare event. 

\begin{table}\caption{\label{tab3}Long-time diffusion coefficients $D_l$ in units of $\sigma^2/\tau$ with $\tau=\sigma\sqrt{m/k_B T}$ for hard dumbbells with elongation $L^*$ at pressure $P^*$, packing fraction $\eta$, and supersaturation $\beta |\Delta \mu | = 1.0$.}
\begin{ruledtabular}
\begin{tabular}{cccc}
  $L^*$ & $P^*$ &$\eta$ & $D_l\tau/\sigma^2$\\
  0.4	& 34.5 & 0.64	& $1.02\times 10^{-4}$\\
  0.5	& 31.2 & 0.63	& $2.47\times 10^{-4}$\\
  0.8	& 24.8 & 0.61	& $2.78\times 10^{-4}$\\
\end{tabular}
\end{ruledtabular}
\begin{flushleft}
\end{flushleft}
\end{table}

\begin{figure}
 \includegraphics[width=0.4\textwidth]{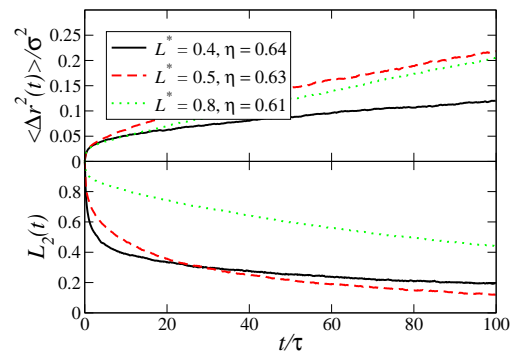}
 \caption{\label{fig7} Mean square displacements $\langle \Delta r^2(t)\rangle$ and $L_2(t)=\langle P_2[\cos(\theta(t))]\rangle$
as a function of time for a fluid of hard dumbbells with an aspect ratio $L^* = 0.4, 0.5$, and $0.8$ at supersaturation $\beta|\Delta \mu| = 1.0$.}
\end{figure}

Additionally, we calculate mean square displacements $\langle \Delta r^2(t)\rangle$ and the second-order orientational correlator $L_2(t)=\langle P_2[\cos(\theta(t))]\rangle$  for a metastable fluid of  hard dumbbells with $L^*=0.4, 0.5$, and   $0.8$ at supersaturation $\beta |\Delta \mu | = 1.0$
as shown in Fig.~\ref{fig7}. We find that at a  supersaturation  $\beta |\Delta \mu | = 1.0$, where the barrier height is still very high, $\Delta G^*/k_B T \sim 170$ for $L^*=0.4$, the long-time diffusion coefficients $D_l\simeq 10^{-4}\sigma^2/\tau$  obtained from $\langle \Delta r^2(t)\rangle$ is extremely small, see Table III), whereas $L_2(t)$ exhibits slow relaxation. Our findings are consistent with predictions obtained from mode-coupling theory for a liquid-glass transition,  in which the structural arrest is due to steric hindrance for both translational and reorientational motion.~\cite{glass1,glass2,glass3,glass4,glass5} Moreover, mode-coupling theory predicts that the steric hindrance  for reorientations becomes stronger with increasing elongation, which is consistent with our results for $L_2(t)$ in Fig.~\ref{fig7}.~\cite{glass1,glass2,glass3,glass4,glass5}. Increasing the supersaturation will lower $\Delta G^*$, but $D_l$ will decrease as well, while at lower supersaturation the barrier height will only increase.
As a result, the nucleation of CP1 phase of hard dumbbells is severely hindered by a high free energy barrier at low supersaturations and slow dynamics at high supersaturations, which  explains why the CP1 phase of colloidal hard dumbbells has never been observed in experiments \cite{imhof2010} or in direct simulations. It is worth noting that the phase diagram might also display (meta)stable CP2 and CP3 close-packed crystal structures,~\cite{vega1992a} which only differ in the way the hexagonally packed dumbbell layers are stacked. As the free energy difference  for the three close-packed structures is extremely small,  we expect  the surface tensions and the nucleation barrier height to be very similar. Hence, we expect that also the nucleation of the CP2 and CP3 phases are hindered by either a high free energy barrier or slow dynamics.

\section{Conclusions}
In conclusion, we  investigated the homogeneous nucleation of the plastic crystal, aperiodic crystal and CP1 crystal phase of hard dumbbells using computer simulations. Hard dumbbells serve as a model system for colloidal dumbbells for which the self-assembly is mainly determined by excluded volume interactions. For charged colloidal dumbbells  or diatomic molecules, screened Coulombic interactions and Van der Waals interactions may significantly change the kinetic pathways for nucleation.   For instance,  crystal nucleation of hard rods proceeds via multi-layered crystalline nuclei whereas attractive depletion interactions between the rods in a polymer solutions   favor the nucleation of single-layered nuclei.~ \cite{ran2010,patti2009} 
For the nucleation of the plastic crystal phase of hard dumbbells, we found
that at low supersaturations the free energy barriers increases slightly with increasing dumbbell anisotropy, which can be explained by a small increase in surface tension for more anisotropic dumbbells.~\cite{mu2006}
When the supersaturation increases, the barrier height decreases with increasing dumbbell aspect ratio, which can only be explained by a different pressure-dependence of the  interfacial tension for hard dumbbells  with different aspect ratios. Although the nucleation rate for the plastic crystal phase  does not vary much with aspect ratio, the dynamics do decrease significantly. We also carried out EDMD simulations and compared the nucleation rates obtained from spontaneous nucleation events with those obtained from the umbrella sampling
Monte Carlo simulations, and found good agreement within the error bars of one order of magnitude.  Additionally, we investigated the structure of the critical nuclei of the plastic crystal phase of hard dumbbells with various aspect ratios. We found that the nuclei of the plastic crystal tend to include more fcc-like particles rather than hcp-like ones, which is similar to the critical nuclei of hard spheres.~\cite{filion2010} However, the amount of hcp-like particles increases with increasing dumbbell aspect ratio, which agrees with the free energy calculations~\cite{marechal2008} where it has been shown that the hcp structure is more stable than fcc structure for  $L^* \ge 0.15$.

Moreover, we also studied the nucleation of the aperiodic crystal phase of hard dumbbells, and our results showed that at the same pressure, the
nucleation barrier of the aperiodic crystal phase of hard dumbbells with $L^*=1.0$ is slightly higher  than that of  hard spheres
which is mostly due to a small difference in supersaturation $\beta|\Delta \mu^{sph}|$. We also performed EDMD simulations for the spontaneous nucleation
of the aperiodic crystal from hard-dumbbell fluid phase, and we found that the nucleation rate obtained from spontaneous nucleation agrees very well
with the one obtained from umbrella sampling MC simulations. Furthermore, we studied the effect of aspect ratio on the nucleation of the aperiodic crystal phase, and found that
at the same supersaturation, the nucleation rate in units of long-time diffusion coefficients
increases for shorter hard dumbbells. However, when the aspect ratio of dumbbells
decreases, the pressure range where the aperiodic crystal phase is stable becomes smaller. Additionally, we also found that the structure of the critical
nuclei of the aperiodic crystal phase formed by hard dumbbells with $L^*=1.0$ is very similar to that of  hard spheres which tend to have
 more fcc-like particles rather than hcp-like ones.

We estimated the height of the free energy barrier for the nucleation of the CP1 crystal phase of hard dumbbells according to classical nucleation theory, which turns out to be extremely high in the normal pressure range due to a high interfacial  tension. Furthermore, we calculated the long-time diffusion coefficients for hard dumbbells at a moderate supersaturation, i.e. $\beta |\Delta \mu| = 1.0$, which appears to be very small.  As a result, we conclude that the high free energy barrier  as well as the slow dynamics  suppress significantly the nucleation of CP1 phase. 

\begin{acknowledgments}
  We thank Matthieu Marechal for offering the equation of state for the crystal structures of hard dumbbell particles and fruitful discussions. Financial support of a NWO-VICI grant is acknowledged.
\end{acknowledgments}


\end{document}